\documentclass{ws-ijmpd}

\begin{document}

\markboth{A. Baushev \& P. Chardonnet}
{Electric charge estimation of a new-born black hole}

%
\catchline{}{}{}{}{}
%

\title{ELECTRIC CHARGE ESTIMATION OF A NEW-BORN BLACK HOLE}

\author{ANTON BAUSHEV}

\address{Bogoliubov Laboratory of Theoretical Physics, Joint Institute for Nuclear Research\\
141980 Dubna, Moscow Region, Russia\\
Universit\'e de Savoie, Laboratoire d'Annecy-le-Vieux de Physique Th\'eorique (LAPTH),\\
BP 110 F-74941, Annecy-le-Vieux Cedex, France\\
abaushev@gmail.com}

\author{PASCAL CHARDONNET}

\address{Universit\'e de Savoie, Laboratoire d'Annecy-le-Vieux de Physique Th\'eorique (LAPTH),\\
BP 110 F-74941, Annecy-le-Vieux Cedex, France\\
chardonnet@lapp.in2p3.fr}

\maketitle

\begin{history}
\received{Day Month Year}
\revised{Day Month Year}
\comby{Managing Editor}
\end{history}

\begin{abstract}
Though a black hole can theoretically possess a very big charge ($Q/(\sqrt{G} M) \simeq 1$), the
charge of the real astrophysical black holes is usually considered to be negligible. This
supposition is based on the fact that an astrophysical black hole is always surrounded by some
plasma, which is a very good conductor. However, it disregards that the black holes have usually
some angular momentum, which can be interpreted as its rotation of a sort. If in the plasma
surrounding the hole there is some magnetic field, it leads to the electric field creation and,
consequently, to the charge separation.

In this article we estimate the upper limit of the electric charge of stellar mass astrophysical
black holes. We have considered a new black hole formation process and shown that the charge of a
new-born black hole can be significant ($\sim 10^{13}$~{Coulombs}). Though the obtained charge of
an astrophysical black hole is big, the charge to mass ratio is small $Q/(\sqrt{G} M) \sim
10^{-7}$, and it is not enough to affect significantly either the gravitational field of the star
or the dynamics of its collapse.
\end{abstract}

\keywords{Black Hole; Electric Charge; Classical Electrodynamics.}

\section{Introduction}
The prediction  and the discovery of stellar mass black holes certainly were a great achievement of the last century.
However, it is still unclear if astrophysical black holes can possess a significant electric charge. The main objection
is related to common observation that there is no charge in the standard astrophysical objects like stars and planets.
We know anyway that all these objects have an electric charge which is nonzero. As an example, the sun charge computed
following Rosseland (1924) is equivalent to a few hundred of coulombs. However, from the point of view  of relativistic
astrophysics, it is insignificant. There is also a theoretical objection\cite{eardley1975}, related to the fact that a
compact object with a net charge greater than $\sim 100 \times (M/M_\odot) \; C$, would be rapidly
discharged by interaction with the surrounding plasma.  \\
The critic remarks, we can address to the two objections, are the following: \\
1) Regarding the observations, we know that it is related to  criteria and precision of the
experiment. In fact, new space-based  missions will be able to measure the position of stars with a
precision of $1 \; \mu as$. This will be sufficient to scrutinize the possible gravitational
lensing and evidence the part due to the  charge of the black hole.\\
2) From the theoretical point of view, we know that the Reissner-Nordstr\"om metric allows a maximum charge given by :
$ Q_{max}= \sqrt {G}\times M \simeq 1.7 \cdot 10^{20} (M/M_\odot) \; C$. We want to  stress that the
objection\cite{eardley1975} is not a theoretical limit since it depends on the modelization of the local  environment
and presupposes that this charge is static. We cannot exclude a peculiar situation when we could dynamically form a
charged  black hole that could evolve after formation to a complete or partial  discharge. Here the key  assumption is
related to a dynamical approach. Without this fence of a static charge for an astrophysical object we can explore how a
compact object could acquire a net charge. In this article, we propose an astrophysical scenario  in order to  obtain a
charged black hole. Our basis is that we have observed numerous compact objects with a high magnetic field ($B \simeq
10^{13} \; Gs$) and with high rotation velocity  ($\Omega \simeq 10^{3} \;  rad.s^{-1}$). The aim of this article is to
show that there is no reason to believe that the charge of a new-born black hole is negligible.

\section{Model description}
As it is believed, stellar black holes are born most likely as a result of supernova explosions of
massive stars\cite{knigashefa2}. At the later stages of evolution this sort of a star consists of a
compact and dense core and an extensive rare envelope. Finally, the core loses stability and
collapses into a black hole. The envelope is erupted out and observed as a supernova
phenomenon\footnote{Strictly speaking, there are other models of the stellar black hole creation,
for instance, with no supernova explosion (so-called silent collapse\cite{knigashefa2}).}.

During all the process the collapsing core is surrounded by some plasma which is a very good
conductor. Usually, it is considered as a critical argument that the core and, consequently, the
new-born black hole cannot have any significant charge. However, it disregards that the core
rotates and possesses a strong magnetic field. If the resistance of the substance is negligible,
the electric field in the comoving frame of reference must be zero. In the static frame of
reference the electric field is defined by the Lorentz transform:
\begin{align}
\vec E &=-\dfrac{\left[\dfrac{\vec \upsilon}{c}\times \vec B\right]}{\sqrt{1-\dfrac{\upsilon^2}{c^2}}}\label{lifsh1} \\
\notag\\
\vec E &=-\left[\dfrac{\vec \upsilon}{c}\times \vec B\right]\qquad \text{in the nonrelativistic limit}\label{lifsh2}
\end{align}
Generally speaking, the charge density $\rho\sim\mathop{\mathrm{div}}\vec E$ is nonzero.
Consequently, the core charge need not be obligatory equal to zero.

In order to estimate the charge of the core we use the following
procedure:
\begin{enumerate}
\item Magnetic field calculation around the collapsing core. \item Determination of the surface
velocity and electric field calculation in accordance with (\ref{lifsh1}). \item The charge
calculation from the Maxwell equation
\begin{equation}
Q=\dfrac{1}{4\pi}\oint \vec E\, d\!\vec S \label{maxwell}
\end{equation}
\end{enumerate}
We make several assumptions to simplify the solution. First of all, the core and the substance
surrounding the core are supposed to have infinite conductivity (see substantiating calculations in
the {\it Discussion}). We presume the gravitational field around the core to be the Schwarzschild
one and the collapse to be spherically symmetric. Of course, for the rotating core it cannot be
precisely correct. However, if the rotation is not very rapid, the deviation from the Schwarzschild
metric should not be very big. Acceptability of the supposition will be discussed in the {\it
Discussion}. We presume that the magnetic field has only the dipole component during the collapse
and the axis of rotation coincides with the axis of the magnetic dipole. We make also some more
assumptions, but they will be mentioned during the solution itself.

\section{Calculations}
Hereafter $t$, $r$, $\theta$, and $\phi$ are the standard Schwarzschild coordinates, $r_g \equiv
\dfrac{2 G M }{c^2}$. We use the orthonormal tetrad frame
 \begin{equation}
  \left( \vec e_0= \dfrac{1}{\sqrt{1-r_g/r}}\;\partial_t,\quad \vec e_1=
\sqrt{1-\dfrac{r_g}{r}}\;\partial_r,\quad \vec e_2= \dfrac{1}{r}\;\partial_\theta,\quad \vec e_3=
\dfrac{1}{r\sin\theta}\;\partial_\phi \right) \label{frame}
\end{equation}
The corresponding coframe of differential forms looks like:
 \begin{equation}
  \left(\omega^0=\sqrt{1-\dfrac{r_g}{r}}\;dt,\quad \omega^1= \dfrac{dr}{\sqrt{1-r_g/r}},\quad  \omega^2=
r\;d\theta,\quad \omega^3= r\sin\theta\;d\phi \right) \nonumber
\end{equation}
With this choice of the frame the metric has the Lorentz form $g_{\alpha\beta}\equiv\eta_{\alpha\beta}$.

 The quantities $\vec B$, $\vec E$, $\vec F$, and $\vec\upsilon$ are three-vectors; their
components are given in the spatial orthonormal triad set of basis three-vectors, corresponding to the
chart~(\ref{frame}).
 $$
  \left(\vec e_r=
\sqrt{1-\dfrac{r_g}{r}}\;\partial_r,\quad \vec e_\theta= \dfrac{1}{r}\;\partial_\theta,\quad \vec
e_\phi= \dfrac{1}{r\sin\theta}\;\partial_\phi \right)
$$

\subsection{Magnetic field calculation}
The dipole magnetic field of a spherically symmetric massive body is described by the
formulae\cite{ginzburg65,anderson70}:
\begin{align}
B_r &=A \left[ -\ln\left(1- \dfrac{r_g}{r}\right)-\dfrac{r_g}{r} -\dfrac{r^2_g}{2 r^2} \right]\cos\theta \label{magn1} \\
B_\theta &=A \left[ \left(\dfrac{r}{r_g}-1\right)\ln\left(1- \dfrac{r_g}{r}\right)+1-\dfrac{r_g}{2
r} \right] \sqrt{\frac{r^2_g}{r(r-r_g)}}\cdot \sin\theta \label{magn2}
\end{align}
The equations are valid for all the space out of the core, but only the surface of the collapsing
core is the object of our interest, and we mean by $r$ hereafter the radius of the core. $A$ is a
coefficient proportional to the dipole momentum of the system.

In order to obtain the magnetic field around the collapsing core we assume\cite{anderson70} the
collapse to be quasistatic . Of course, this supposition is artificial, especially of the last
stages of collapse, but it gives us the opportunity to proceed a strict examination of the problem
in the framework of general relativity.

If the collapse is quasistatic, the magnetic field around the core is always described by
(\ref{magn1},\ref{magn2}) but the coefficient $A$ is no longer constant\cite{anderson70}, it
depends upon the radius of the collapsing core $r$. We can deduce the dependence from infinite
conductivity of the star and continuity of the normal component of the magnetic field on the
surface of the core. In our case it can be written in the form (see Ref.~\refcite{anderson70},
equation(24) for more details): $B_r  r^2 \sin^2\theta=\mbox{\it const}$. Since the collapse is
radial, $\theta$ coordinate of a surface element considered remains constant during it and,
consequently, ($B_r r^2$) and ($B_r r^2 / \cos\theta$) are also constant. We define a new quantity
$D\equiv B_r  \dfrac{r^2}{r^2_g \cos\theta}$  which thus remains constant during the collapse (in
contrast to $A$). It should be derived from the initial conditions. We obtain from (\ref{magn1}):
\begin{align}
&A\left[ -\dfrac{r^2}{r^2_g}\ln\left(1- \dfrac{r_g}{r}\right)-\dfrac{r}{r_g} -\dfrac12 \right]= D\\
&A=D\left[ -\dfrac{r^2}{r^2_g}\ln\left(1- \dfrac{r_g}{r}\right)-\dfrac{r}{r_g} -\dfrac12
\right]^{-1} \label{a}
\end{align}
Substituting this $A$ into (\ref{magn1}) and (\ref{magn2}), we get the formulas describing the
magnetic field around the collapsing core:

\begin{align}
B_r &= D\dfrac{r^2_g}{r^2}\cos\theta \\
B_\theta &=D \; \sin\theta\; \sqrt{\frac{r^2_g}{r(r-r_g)}} \left(
\dfrac{\left(\dfrac{r}{r_g}-1\right)\ln\left(1- \dfrac{r_g}{r}\right)+1 -\dfrac{r_g}{2 r}}{
-\left(\dfrac{r}{r_g}\right)^2 \ln\left(1- \dfrac{r_g}{r}\right)-\dfrac{r}{r_g} -\dfrac12}\right)
\label{magth}
\end{align}

\subsection{Calculation of the charge created by the collapsing core rotation in the magnetic
field}

Tangential velocity of a particle falling on a black hole  is defined by the
angular momentum conservation condition\cite{teorpol}. For a non-relativistic particle falling in the equatorial plane with small angular momentum
$\mathfrak{M}$ we have:
\begin{equation}
\upsilon_{\phi} = \dfrac{\alpha c}{r}\sqrt{\dfrac{r-r_g}{r}} \nonumber
\end{equation}
Here $\alpha\equiv\dfrac{\mathfrak{M}}{mc}$ is a constant. Since the angular momentum of the
collapsing core also remains constant, by analogy we can suggest the tangential velocity of the
surface to be
\begin{equation}
\upsilon_{\phi} = \dfrac{\alpha c}{r}\sqrt{\dfrac{r-r_g}{r}}\: \sin\theta \label{tangen}
\end{equation}
Here $\sin\theta$ provides solid body rotation of the surface, and $\alpha$ is as before a
constant which should be derived from the initial conditions.

In order to calculate electric field intensity, we should make some assumptions about the
substance surrounding the collapsing core. It is most likely hot but respectively rare plasma. As
it was already mentioned, we consider it as a perfect conductor. Since the magnetic field of the
core is very strong at the late stages of the collapse, we can expect that the plasma is frozen to
the magnetic field and corotates. Consequently, its tangential velocity $\upsilon_{\phi}$ (just
around the surface of the core) is equal to (\ref{tangen}). The question of $\upsilon_{r}$ and
$\upsilon_{\theta}$ components of the plasma velocity is not so clear. We speculate that they are
equal to zero. Properly speaking, in the nonrelativistic case they cannot make a contribution to
the electric field flux (since $B_{\phi}=0$), but if they are relativistic, they contributes to
the $\upsilon^2$ in (\ref{lifsh1}). Hence, in our case the supposition $\upsilon_{r}=0$,
$\upsilon_{\theta}=0$ means only that the velocities are much less than the speed of light.

Finally, for the substance velocity we have:
\begin{equation}
\upsilon_{r}=0 \quad \upsilon_{\theta}=0 \quad \upsilon_{\phi} = \dfrac{\alpha
c}{r}\sqrt{\dfrac{r-r_g}{r}}\: \sin\theta \label{velocity}
\end{equation}

Since we presume the rotation of the core to be slow, the ratio $\alpha/r_g$ is small and
$\upsilon_{\phi} \ll c$; in any case near the gravitational radius $\upsilon_{\phi} \to 0$. We can
therefore use (\ref{lifsh2}) to calculate the electric field. Combining (\ref{magth}) and
(\ref{velocity}), we obtain for the electric field just above the core surface:
\begin{equation}
E_r =\frac{\alpha D r_g}{r^2} \left( \dfrac{\left(\dfrac{r}{r_g}-1\right)\ln\left(1-
\dfrac{r_g}{r}\right)+1 -\dfrac{r_g}{2 r}}{ -\left(\dfrac{r}{r_g}\right)^2 \ln\left(1-
\dfrac{r_g}{r}\right)-\dfrac{r}{r_g} -\dfrac12}\right) \sin^2\theta \nonumber
\end{equation}
Integrating it over the sphere surrounding the core and using equation (\ref{maxwell}), we obtain
for the charge:
\begin{equation}
Q =\frac{\alpha D r_g}{3 \pi} \left( \dfrac{\left(\dfrac{r}{r_g}-1\right)\ln\left(1-
\dfrac{r_g}{r}\right)+1 -\dfrac{r_g}{2 r}}{ -\left(\dfrac{r}{r_g}\right)^2 \ln\left(1-
\dfrac{r_g}{r}\right)-\dfrac{r}{r_g} -\dfrac12}\right) \label{charge}
\end{equation}
The charge is positive if the angular velocity of the core and its magnetic moment are
codirectional.

\section{Discharge processes and the new-born black hole charge estimation}
The charge described by (\ref{charge}), reaches its maximum at $r\simeq 1.3 r_g$ and drops to zero
as $|\ln((r-r_g)/r_g)|^{-1}$ when $r$ tends to $r_g$. Seemingly, it should mean that the charge of
the formed black hole is zero. However, we have not yet taken into account gravitational force and
its influence on the discharge processes. Let us consider a resting probe particle of mass $m$ and
charge $q$ (let us denote its specific charge by $\mu$). Electric field acting on the particle
consists of several components: electric field (\ref{lifsh1}) produced by the plasma rotation in
the magnetic field and electrostatic field created by the excess of nonequilibrium charge (with
respect to (\ref{charge})) which has not discharged. Magnetic field impedes the discharge
(equilibrium quantity of charge, described by (\ref{charge}), is always nonzero). However, since
the equilibrium charge approaches zero when $r\to r_g$, we can ignore for simplicity the influence
of the magnetic field at the last stages of the collapse.

\begin{figure}
\centerline{\epsfig{file=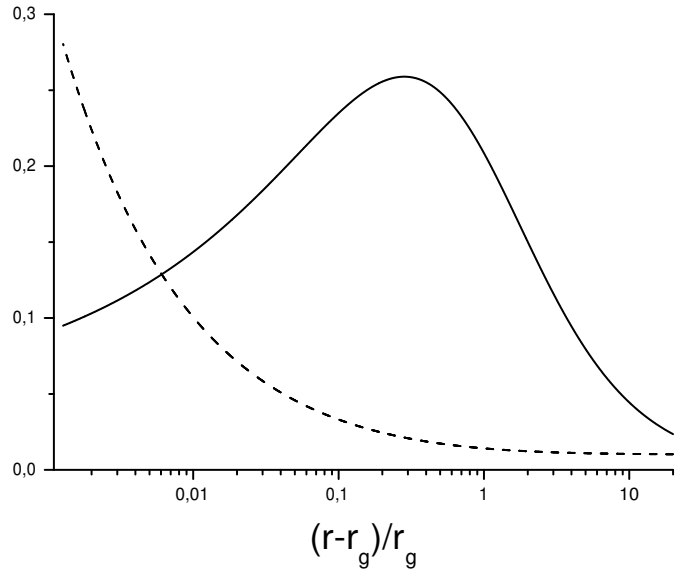, width=10cm}} \caption{represents the left (solid line) and
the right (dot line) parts of equation (\ref{main}) for the case when $\mathfrak{Y}=0.01$. They
are proportional to the charge created by the collapsing core rotation in the magnetic field and
to the charge which can be held by the gravitational field, respectively, with the coefficient
depending upon the parameters of the system.} \label{fig}
\end{figure}

The electrostatic force acting on the probe particle can be taken as:
\begin{equation}
F_e=\dfrac{q Q}{r^2}
\end{equation}
Gravitational force acting on the particle is\cite{frolovnovikov,zeldnov1}:
\begin{equation}
F_g=-\dfrac{G M m}{r^2 \sqrt{1-\dfrac{r_g}{r}}}
\end{equation}
Both the forces directed radially. The discharge (by particles with specific charge $\mu$) will be
stopped by the gravitational field, when $F_e+F_g=0$.
\begin{align}
\dfrac{q Q}{r^2}&=\dfrac{G M m}{r^2 \sqrt{1-\dfrac{r_g}{r}}} \nonumber\\
Q&=\dfrac{r_g c^2}{2\mu}\sqrt{\dfrac{r}{r-r_g}} \label{gcharge}
\end{align}
The last equation represents the charge which can be held by the gravitational field. It is
represented on {\it Fig.}~\ref{fig} by the dot line)\footnote{Actually, the figure represents the
left (solid line) and the right (dot line) parts of the equation (\ref{main}), but they are
proportional to the charge created by the collapsing core rotation in the magnetic field and to
the charge which can be held by the gravitational field, respectively, with the coefficient
depending upon the parameters of the system.}. The equilibrium charge described by (\ref{charge})
is represented by the solid line. The graphs have the only crossing point. When it is reached, the
discharge is stopped by the gravitational field. Consequently, we can take the charge of the
collapsar at this point as an estimation of the charge of new-born black hole. We equate
expressions (\ref{charge}) and (\ref{gcharge}):
\begin{equation}
\frac{\alpha D r_g}{3 \pi} \left( \dfrac{\left(\dfrac{r}{r_g}-1\right)\ln\left(1-
\dfrac{r_g}{r}\right)+1 -\dfrac{r_g}{2 r}}{ -\left(\dfrac{r}{r_g}\right)^2 \ln\left(1-
\dfrac{r_g}{r}\right)-\dfrac{r}{r_g} -\dfrac12}\right) = \dfrac{r_g
c^2}{2\mu}\sqrt{\dfrac{r}{r-r_g}}\nonumber
\end{equation}
Introducing a new dimensionless constant
\begin{equation}
\mathfrak{Y}=\dfrac{3\pi c^2}{2\alpha \mu D} \label{y}
\end{equation}
we finally obtain:
\begin{equation}
\left( \dfrac{\left(\dfrac{r}{r_g}-1\right)\ln\left(1- \dfrac{r_g}{r}\right)+1 -\dfrac{r_g}{2 r}}{
-\left(\dfrac{r}{r_g}\right)^2 \ln\left(1- \dfrac{r_g}{r}\right)-\dfrac{r}{r_g}
-\dfrac12}\right)=\mathfrak{Y}\sqrt{\dfrac{r}{r-r_g}} \label{main}
\end{equation}
In order to find the charge we should solve this equation and substitute the obtained radius $r$ into (\ref{gcharge})
or (\ref{charge}). If $\mathfrak{Y}$ is very small ($-\ln\mathfrak{Y}\gg 1$), then $-\ln(r/r_g-1)\gg 1$, and the
equation can be simplified:
\begin{equation}
\dfrac{1}{2 \mathfrak{Y}}=\sqrt{\frac{r_g}{r-r_g}}\ln\frac{r_g}{r-r_g} \label{mainsimple}
\end{equation}

\section{Application of the results for calculation of the charge of an astronomical black hole}
Let us estimate the electric charge of a real new-born black hole. We have already mentioned that
a sun mass black hole is born by the collapse of a massive star central part. As it is believed,
neutron stars are also born as a consequence of massive star core collapses, but of lower masses.
Consequently, the characteristic physical parameters of the core can be estimated from the
parameters of neutron stars. The radius of a neutron star is $10-15$~km, the mass is $1.4-1.8
M_\odot$. Pulsars have the magnetic field $10^{12}-10^{13}$~{Gs}, the shortest known period of
revolution\cite{knigashefa1} of a pulsar is $1.5$~ms.

We take the mass of the collapsing core $3 M_\odot$ ($r_g=9\cdot 10^5$~cm). It is heavier than a
neutron star to form a black hole. Hence, we should consider its physical parameter when its
radius is bigger. Using the above-mentioned analogy, we presume that when the radius of the
collapsing core is $20$~{km} its period of revolution is $10^{-3}$~{s} and the magnetic field
intensity on its pole is $10^{13}$~{Gs}. As the specific charge $\mu$ we take the specific charge
of electron (or positron). If the charge $Q$ is negative (the angular velocity of the core and its
magnetic moment are oppositely directed), it is obvious. But even in the case when the charge of
the core is positive, the temperature at the last stages of the collapse is so high that the
positron concentration is sufficient to provide the discharge. From this we calculate $\alpha$,$D$
and $\mathfrak{Y}$. The formula for the period of revolution coincides with the classical one:
$T_{\text{\it rev}}=\dfrac{2\pi r}{c \upsilon_{\phi}}$. Subtracting $\upsilon_{\phi}$ from
(\ref{velocity}), we finally find $\alpha\simeq 11$~{km.}

We find $\mathfrak{Y}=1.4\cdot 10^{-16}$. As soon as $-\ln\mathfrak{Y}\simeq 36\gg 1$ we can use
(\ref{mainsimple}). Solving the equation, we have:
\begin{align}
\sqrt{\frac{r-r_g}{r_g}}&=1.77\cdot 10^{-14}\label{rad}\\
Q&=1.4\cdot 10^{13}\: \text{Coulombs} \label{q}
\end{align}
The maximum charge of the core is reached (in accordance with (\ref{charge})), when the radius is
$r\simeq 1.3 r_g$. It is equal to $Q_m=4.3\cdot 10^{14}$~{Coulombs}, thirty times higher than the
final one.

\section{Discussion}
The foregoing calculations may provoke several objections. First of all, the radius (\ref{rad}),
where the discharge is stopped by the gravitational field, is too close to the Schwarzschild
sphere. The disturbance of the sphere produced by the core rotation is much stronger, and the
adduced solution is certainly not valid for this region. Moreover, we used a lot of strong
simplifying suppositions, such as the calculation of the magnetic field on the assumption of
quasistatic collapse. These remarks are correct, but they do not affect significantly the
conclusions of the article. The purpose of the work was to show that a new-born black hole can
possess a big electric charge. The charge of the core surpasses $10^{14}$~C since the radius of
the collapsing core becomes less than $r\simeq 8 r_g$, where the departure of the metric from the
Schwarzschild one is not so significant. Moreover, the final charge depends weakly on the
parameter (\ref{rad}) and, consequently, on the details of the discharge processes. Of course, the
estimation (\ref{q}) can be inaccurate, but nevertheless we have strong reasons to presume the
final charge to be bigger than $10^{13}$~C.

Initially, the star is not electrically charged. As we have shown, because of the magnetodynamical
processes a charge separation appears. Then what happens with the charge separated from the
forming black hole? Of course, it is emitted to the substance of the envelope; a further destiny
of the charge depends on the envelope evolution. The collapse of a core of a massive star is
considered numerically in Ref.~\refcite{ardeljan05}. It is shown that the differential rotation of
the envelope appears resulting in the magnetic field generation. At some moment
magnetohydrodynamical instability arises; it leads to the twisting of the magnetic lines and
closed magnetic vortex generation. The magnetic field structure becomes very complex. Finally,
magnetic pressure increases so that it ejects the envelope to the outer space producing a
supernova explosion. The charge evolved from the core moves along the magnetic lines and is
trapped by the forming vortexes. During the supernova explosion the envelope substance is erupted
out together with the charge frozen in the magnetic vortexes.

In the {\it model description} we presumed that the core and the substance surrounding the core
have infinitive conductivity. This statement can be proved numerically. Conductivity of the
totally ionized plasma depends only on its temperature and does not depend on the
density\cite{knigashefa1}:
\begin{equation}
\sigma=8\cdot 10^{-4} T^\frac32 _e \left[\frac{1}{\text{Ohm}\cdot \text{m}}\right]\nonumber
\end{equation}
Let us consider a sphere of the radius of $20$~{km} (characteristic for the collapsing core) and
with electric charge $1.4\cdot 10^{13}$~C merged into the plasma with the temperature
$T_e=10^{6}$~K. Then electric field intensity around the sphere is $6\cdot
10^{14}$~$\text{V}/\text{m}$, current density is $5\cdot 10^{20}\text{A}/\text{m}^2$, and the
total current is $2.5\cdot 10^{30}\text{A}$. The sphere would be discharged in $\sim 10^{-17}$~s.
Of course, this time has not a direct physical meaning, but this proves the negligibility of the
electrical resistance in the considered task.

It is important to notice that though the obtained charge of an astrophysical black hole is big,
the charge to mass ratio is small $Q/(\sqrt{G} M) \sim 10^{-7}$, and it is not sufficient to
affect significantly either the gravitational field of the star or the dynamics of its collapse.
The  numerical collapse of a charged star in the Reissner-Nordstr\"om space-time was
computed\cite{ghezzi2004,ghezzi2005}. They considered spherical symmetry with distribution of an
electric charge proportional to the distribution of mass. For the  numerical computation they
assumed a polytropic equation of state with $\gamma = 5/3$.   They concluded that for low values
of $Q/(\sqrt{G} M) $ no departure from  unpolarized collapse was found. In our case $Q/(\sqrt{G}
M) \sim 10^{-7}$, which satisfies the criteria\cite{ghezzi2004}, and it also justifies our
approach of using a Schwarzschild metric.

In the article, we considered a collapsing core with a moderate magnetic field. The electric
charge of the new-born black hole turns to be sizable, but its electrostatic field ($\sim
10^{12}$~{Gs}) is lower than the critical one ($4.4\cdot 10^{13}$~{Gs}). However, in the case of
magnetars, the magnetic field of the collapsing core can also be extremely strong ($\sim
10^{15}$~{Gs}). Then the charge of the black hole can be so big that the electric field will
surpass the critical. In this case a new effect appears\cite{preparata98,ruffini2001}: a very
large number of $e^+ e^-$ pairs is created in the region with overcritical electric field, which
finally leads to the formation of a strong gamma-ray burst.

Interaction of the accreting substance with the energetic $\gamma$-ray emission of a black hole
can also induce an electric charge into it\cite{diego2004}. The Compton scattering of
$\gamma$-photons kicks out electrons from the accretion flow, while for the protons this mechanism
is not so effective because of higher mass-charge ratio. As a result, the black hole should obtain
a significant charge. However, one can see\cite{diego2004} that in the system considered the
charge induced by this process is approximately 100 times lower than (\ref{q}). Actually, in our
case the mechanism is not effective.

In Ref.~\refcite{lee2001}, processes in the force-free magnetosphere of an already formed black
hole are considered. As it was shown, because of accretion the black hole acquires a significant
electric charge:
\begin{equation}
Q\sim 10^{15}\text{C} \left(\dfrac{B}{10^{15}\text{Gs}}\right)\left(\dfrac{M}{M_\odot}\right)^2 \nonumber
\end{equation}
Here $B$ is the magnetic field intensity in the accreting substance near the black hole. This
quantity is big and quite correlates with (\ref{q}). Thus, an astrophysical black hole can possess
a big electric charge, either getting it during the birth or owing to the accretion.

\section{Acknowledgements} This work was supported by the RFBR (Russian Foundation for Basic
Research, Grant 08-02-00856).


\begin{thebibliography}{99}

\bibitem{eardley1975}
D. M. Eardley and W. H. Press, (1975), {\it Annual review of astronomy and astrophysics}, {\bf 13}, 381

\bibitem{knigashefa2}
G. S. Bisnovatyi-Kogan, (2002)
{\it Stellar Physics 2: Stellar Evolution and Stability},
Series: Astronomy and Astrophysics Library, Springer.

\bibitem{ginzburg65}
V. L. Ginzburg and L. M. Ozernoi, (1965), {\it Soviet Phys. JETP}, {\bf 20}, 689

\bibitem{anderson70}
J. L. Anderson and J. M. Cohen, (1970), {\it Ap\&SS}, {\bf 9}, 146

\bibitem{teorpol}
L. D. Landau and E. M. Lifshitz,  {\it The Classical Theory of Fields}, {\it any edition}

\bibitem{knigashefa1}
G. S. Bisnovatyi-Kogan, (2001)
{\it Stellar Physics 1: Stellar Evolution and Stability},
Series: Astronomy and Astrophysics Library, Springer.

\bibitem{ardeljan05}
N. V. Ardeljan, G. S. Bisnovatyi-Kogan, S. G. Moiseenko, (2005), {\it MNRAS}, {\bf 359}, 333

\bibitem{frolovnovikov}
V. P. Frolov, I. D. Novikov,  {\it Black hole physics. Basic concepts and new developments}, {\it
Kluwer}, (1997)

\bibitem{zeldnov1}
Ya. B. Zel'dovich,  I. D. Novikov,  {\it Relativistic Astrophysics, Vol. 1: Stars and Relativity.}
Mineola, NY: Dover Publications, (1996)

\bibitem{ghezzi2004}
C.R. Ghezzi, P.S. Letelier, astro-ph/0503629

\bibitem{ghezzi2005}
C.R. Ghezzi, (2005), {\it Physical Review D}, {\bf 72}, Issue 10, 104017

\bibitem{preparata98}
G. Preparata, R. Ruffini, S.-S. Xue, (1998), {\it A\&A}, {\bf 338}, L87.

\bibitem{ruffini2001}
R. Ruffini, C. Bianco, F. Fraschetti, et.al, (2001), {\it Ap.J. Letters}, {\bf 555}, 107

\bibitem{diego2004}
J.A. de Diego, D. Dultzin-Hacyan, J.G. Trejo, D. N\'u\~nez, (2004),  astro-ph/0405237

\bibitem{lee2001}
H.K. Lee, C.H. Lee, M.H.P.M. van Putten, (2001), {\it MNRAS}, Volume 324, Issue 3, pp. 781-784.

\end{thebibliography}
\end{document}